                                                                  \documentclass[12pt,preprint]{aastex}

\usepackage{lineno}
\shorttitle{splitting and eruption of an active region filament}
\shortauthors{Kong et al.}
\linenumbers
\begin{document}

\nolinenumbers

\title{Splitting and eruption of an active region filament caused by magnetic reconnection}
\author{Defang Kong\altaffilmark{1,2}, Jincheng Wang\altaffilmark{1,2}, Genmei Pan\altaffilmark{3}}

\altaffiltext{1}{Yunnan Observatories, Chinese Academy of Sciences, Kunming 650216, People's Republic of China; kdf@ynao.ac.cn}
\altaffiltext{2}{Yunnan Key Laboratory of Solar Physics and Space Science, Kunming 650216, People's Republic of China}
\altaffiltext{3}{College of Data Sciences, Jiaxing University, Jiaxing 314001, People's Republic of China}

\begin{abstract}

To gain a deeper understanding of the intricate process of filament eruption, we present a case study of a filament splitting and eruption by using multi-wavelength data of the Solar Dynamics Observatory (SDO). It is found that the magnetic reconnection between the filament and the surrounding magnetic loops resulted in the formation of two new filaments, which erupted successively. The observational evidences of magnetic reconnection, such as the obvious brightening at the junction of two different magnetic structures, the appearance of a bidirectional jet, and subsequent filament splitting, were clearly observed. Even though the two newly formed filaments experienced failed eruptions, three obvious dimmings were observed at the footpoints of the filaments during their eruptions. Based on these observations, it is suggested that magnetic reconnection is the trigger mechanism for the splitting of the original filament and the subsequent eruption of the newly formed filaments. Furthermore, the process of filament splitting dominated by magnetic reconnection can shed light on the explanation of double-deck filament formation.

\end{abstract}

\keywords{Sun: activity - Sun: filaments - Sun: prominences - Sun: magnetic fields - Sun: magnetic reconnection}

\section{Introduction}

Solar filaments are elongated dark structures against the bright solar disk and can be observed in absorption in H$\alpha$ and extreme ultraviolet (EUV) channels, with relatively cold and dense plasmas suspended in the extremely hot solar corona (Rust \& Kumar 1994; Tandberg-Hanssen 1995; Demoulin 1998; Gilbert, Alexander \& Liu 2007). If observed above the solar limb, they can be seen in emission and are called prominences. Observations indicate that they often form along polarity inversion lines (PILs) in filament channels between the regions with opposite magnetic polarities (Mackay et al. 1997; Gaizauskas et al. 1997; Martin 1998; Gaizauskas 1998).

Due to the close relationship with solar flares and coronal mass ejections (CMEs), solar filaments have continuously attracted intensive attention (Smith, Hildner \& Kuin 1992; Jing et al. 2004, Yan \& Qu 2007). Observations indicate that the shear motion and the convergence of magnetic fields with opposite polarities is a necessary condition for the formation of filaments (van Ballegooijen \& Martens 1989; Martin 1990; Gaizauskas 1998; Yan et al. 2020b). Magnetic flux cancellation is also considered to play an important role in the formation of filaments (Martin, Livi \& Wang 1985; Martin 1998; Wang \& Muglach 2007; Yan et al. 2016), sustaining newly formed filaments (Mackay et al. 2010) and modifying the magnetic field configuration of filaments, so as to trigger eruptions (Zuccarello et al. 2007; van Driel-Gesztelyi et al. 2014). Martens and Zwaan (2001) presented a model for the formation, evolution and eruption of filaments, on the basis of the photospheric magnetic flux convergence and cancellation. 

A stable filament could erupt due to losing the balance of forces acting on it. The investigation on the triggering mechanism of filament eruption is a prominent issue in solar physics. Several triggering mechanisms have been proposed to explain observed solar eruptions, such as the flux emergence model (Chen \& Shibata 2000; Wang et al. 2016), the flux cancellation model (Livi et al. 1989; Linker et al. 2003; Sterling et al. 2018), the helical kink instability model (T{\"o}r{\"o}k, Kliem \& Titov 2004; Fan 2005; T{\"o}r{\"o}k \& Kliem 2005), the torus instability model (Kliem \& T{\"o}r{\"o}k 2006; Schrijver et al. 2008; Fan 2010), the breakout model (Antiochos et al. 1999), the magnetic catastrophe model (Forbes \& Isenberg 1991; Lin \& Forbes 2000), the tether-cutting model (Moore, Sterling \& Hudson 2001), and so on. These models can be divided into two types: one is ideal magnetohydrodynamic (MHD) process, the other is non-ideal MHD process.

Solar eruptions are generally thought to involve magnetic reconnection, a process in which the efficient release of magnetic energy can give rise to plasma heating and particle acceleration in a narrow region (Coppi \& Friendland 1971; Lin \& Forbes 2000; Xue et al. 2016; Yan et al. 2022). Both observations and models propose that there are two kinds of magnetic reconnections (Sterling \& Moore 2004), i.e., internal magnetic reconnection (IMR) (Gilbert et al. 2001; Gibson \& Fan 2006; Tripathi et al. 2009; Zhang et al. 2022) and external magnetic reconnection (EMR) (Zhou et al. 2017), which largely depend on the proposed pre-eruption magnetic topology. In filament eruptions, magnetic reconnection between two groups of magnetic structures below and above the filaments is found to play a significant role (Forbes \& Isenberg 1991; Antiochos, Devore \& Klimchuk 1999; Moore at al. 2001; Sterling \& Moore 2004; Schrijver 2009; Chen et al. 2016). Besides, the interaction between the flux rope and the surrounding pre-existent magnetic system also seems to play an important role in triggering solar eruptions. When positive and negative helicities coexist in a single domain, two magnetic flux systems with opposite magnetic helicity signs can merge via magnetic reconnection, resulting in the decrease and cancellation of the total magnetic helicity in the active region (Romano et al. 2011a). By measuring the magnetic helicity flux in the corona, Romano et al. (2011b) found that filament eruptions may be caused by magnetic reconnection between two magnetic field systems with opposite signs of magnetic helicity. Additionally, Romano et al. (2014) proposed that the magnetic flux and magnetic helicity are only partial indicators of the real likelihood of an eruptive event. Usually, observations indicate that some filaments are split during their eruptions (Contarino et al. 2003; Liu, Alexander \& Gilbert 2007; Cheng et al. 2018). However, the exact reason for filament splitting remains unclear.

In this study, we analyze observations of filament splitting and subsequent eruption observed by the Solar Dynamics Observatory (SDO; Pesnell et al. 2012) on March 9, 2019. The observations are briefly introduced in Section 2. The results are presented in Section 3. The conclusion and discussion are given in Section 4.

\section{Observations}

The Atmospheric Imaging Assembly (AIA; Lemen et al. 2012) on board SDO can provide simultaneous multiple high resolution full-disk images of the transition region and the corona in seven Extreme Ultraviolet (EUV) and two Ultraviolet (UV) passbands. The spatial resolution of EUV and UV images from SDO/AIA is 0.$^\prime$$^\prime$6 per pixel and the time cadence is 12 s. In this paper, we mainly use EUV imaging data acquired at 304 \AA\ and 211 \AA\ to investigate the evolution process of the filament. 

Additionally, the Helioseismic and Magnetic Imager (HMI; Schou et al. 2012), also on board SDO, can provide full disk line-of-sight magnetograms with a spatial resolution of 0.$^\prime$$^\prime$5 per pixel and a cadence up to 45 s. The SDO/HMI magnetograms are employed to analyze the evolution of the photospheric magnetic fields. Vector magnetograms from Space Weather HMI Active Region Patch (SHARP) series, observed by HMI (Bobra et al. 2014; Centeno et al. 2014), are derived by using the Very Fast Inversion of the Stokes Vector algorithm, with a pixel scale of about 0.$^\prime$$^\prime$5 per pixel and a cadence of 12 minutes (Borrero et al. 2011). The 180-degree azimuthal ambiguity is resolved by using the minimum energy method (Metcalf 1994; Metcalf et al. 2006; Leka et al. 2009). The images are remapped using Lambert (cylindrical equal area) projection, centered on the midpoint of the active region, and tracked at the Carrington rotation rate (Sun 2013).

Moreover, the vector magnetic data obtained by SDO/HMI are used to investigate the magnetic field evolution associated with the evolution of the filament. Taking the photospheric vector magnetic field as the boundary condition, we employed the nonlinear force-free field (NLFFF) to extrapolate the magnetic fields from the photosphere to the corona. We used an optimization algorithm proposed by Wheatland et al. (2000) and implemented by Wiegelmann (2004) to extrapolate the three-dimensional NLFFF structure of active region NOAA 12734. We applied a preprocessing procedure to the bottom boundary vector data before the extrapolation. This procedure aims to eliminate most of the net force and torque which would lead to inconsistencies between the forced photospheric magnetic field and the force-free assumption in the NLFFF models (Wiegelmann et al. 2006). 

The observation time of the magnetogram used for the NLFFF extrapolation is 11:00:09 UT on March 9, 2019, and its resolution is 0.$^\prime$$^\prime$5 pixel$^{-1}$. We used the full-disk vector magnetogram and converted it to Cylindrical Equal Area (CEA) projection coordinates using the ``bvec2cea.pro" routine in the Solar Software Package (SSW). Then, prior to the extrapolation, we performed a 2 $\times$ 2 rebinning of the boundary data which is about 0.72 Mm pixel$^{-1}$. The region selected for the extrapolation span a box of 300 $\times$ 196 $\times$ 196 uniform grid points, corresponding to about 216 $\times$ 141 $\times$ 141 Mm$^3$. The region covers the entire active region of interest, which is essential for the NLFFF model.

All data have been processed up to Level 1.5 by using the standard procedures in the Solar Software Package (SSW) and were differentially rotated to a reference time at 13:10:30 UT on March 9, 2019. 

\section{Results}

NOAA active region (AR) 12734 emerged on the solar disk at about 03:00 UT on March 4, 2019. It gradually evolved into a bi-polar active region. The leading sunspot exhibited a negative polarity , while the following sunspot displayed a positive polarity. During the evolution of this active region, a filament erupted and a B6.1 flare is produced on March 9, 2019. The flare began at 12:17 UT, peaked at 12:26 UT, and ended at 12:36 UT. 

\subsection{Filament activation before its eruption}

Figures 1(a)-(d) show the evolution of the filament in 304 \AA\ wavelength on March 9, 2019. The blue dotted line in Figure 1(a) outlines the initial sigmoid structure of the filament. During the evolution of the filament, a sudden brightening, marked by the blue arrow in Figure 1(b), emerged beneath the filament at about 04:10 UT, indicating the energy release. Subsequently, the entire filament was activated, appearing bright and deformed, with its threads seen winding around each other, as illustrated in Figures 1(c) and 1(d). The eruption process can be observed in Supplementary Movie 1. Figure 1(e) displays the normalized intensity variation of the selected region marked by the black box in Figure 1(d), using 304 \AA\ images. The normalized intensity refers to dividing the  intensity value by the maximum value registered in the time interval in the selected region. It can be clearly seen from Figure 1(e) that an obvious brightening appeared at around 04:10 UT, followed by a higher peak in the intensity profile indicated by the red arrow. Then, the filament became relatively quiet. Until about 12:02 UT, it began to erupt. Related to the eruption, a B6.1 flare occurred. The green arrow indicates the brightening associated with the flare. The vertical blue dotted line marks the onset of the filament eruption. 

The brightening observed beneath the filament is closely linked to the emergence of magnetic fields. Figures 2(a)-(d) display the HMI line-of-sight magnetograms, illustrating the distribution and temporal evolution of magnetic fields with positive and negative polarities highlighted in white and black, respectively. In Figure 2(a), the red box indicates the region of the brightening observed in the 304 \AA\ image at 04:10 UT. Figure 2(e) presents the temporal evolution of the magnetic fluxes in the active region outlined by the red box in Fig. 2(a). The black and blue line indicate the evolution of the positive and negative magnetic fluxes in the selected region, respectively. The positive magnetic fields emerged in this region from 00:00 UT to 04:30 UT as indicated by the black line in Figure 2(e). The negative magnetic fields (the blue line in Figure 2(e)) in the red box gradually decreased from around 03:00 UT, suggesting cancellation with the emerging positive magnetic fields. The magnetic cancellation can be also observed from the evolution of the magnetogram in Supplementary Movie 2. The relative hot materials are observed to propagate from the initial brightening position to the two footpoints of the filament. The whole filament became bigger and brighter than before. The moving hot materials also outline the magnetic structures of the filament. Based on these observations, it is suggested that the filament should be activated by the magnetic reconnection between the emerging magnetic fields and the filament's magnetic structure.

\subsection{Filament splitting, eruption, and dimmings}

After its initial activation, the filament did not erupt immediately. About seven hours later, an obvious brightening appeared again at the eastward part of the filament. This indicates a second activation of the filament, which subsequently became bigger than before. At 11:46:53 UT, the filament exhibited an obvious S-shaped structure. At about 12:02 UT, the eastward part of the filament began to erupt. The eruption process is illustrated using 304 \AA\ and 211 \AA\ images in Figures 3(a)-(f), with the filament outlined by the blue dotted lines in Figures 3(a) and 3(d). At about 12:02 UT, a brightening occurred at the eastward part of the filament as indicated by the blue arrows in Figures 3(b) and 3(e). Then, the filament began to bifurcate, as shown by the blue arrow in Figure 3(c) and the black box in Figure 3(f). Figure 3(g) shows the intensity profile of the selected region outlined by the black box in Figure 3(f) using 211 \AA\ images. The black curve represents the normalized intensity variation relative to the maximum value recorded from 11:00 UT to 15:00 UT on March 9, 2019. The red arrow in Figure 3(g) denotes the emergence of the brightening resulting from the magnetic reconnection between the filament and the surrounding magnetic loops. The subsequent peak illustrates the brightening of the flare loops, observed in all EUV observations. The magnetic reconnection between the filament and the surrounding magnetic loops acquired at 304 \AA\ can be observed in Supplementary Movie 3.

The eruption process of the filament is shown in Figures 4(a)-(f) using 211 \AA\ images. In Figure 4(a), the blue dotted line outlines the shape of the filament before its eruption, while the cyan dotted line outlines the surrounding magnetic loops at the east of the filament. At about 12:02 UT, the magnetic loops reconnected with the filament. The brightening appeared at the joint of the magnetic loops and the filament, as also observed in Figures 3(b)-(f). During its eruption, the westward part of the filament was observed to split into three branches, as evidenced by the three different colored lines in Figure 4(d). In the following, the filament split into two distinct filaments. The blue dotted lines in Figures 4(b)-(f) represent the longer newly formed filament, while the yellow dotted lines in Figures 4(e) and 4(f) indicate the other newly formed filament. In Figure 4(a), the slits AB and CD indicate the positions where the time slices in Figure 6 are made. As far as our knowledge is concerned, this is the first time to observe that the formation of two new filaments is due to the magnetic reconnection between the original filament and the surrounding magnetic loops. The configuration of the two filaments formed a double-decker structures. Therefore, magnetic reconnection may be an important candidate mechanism for the formation of double-decker filament. The splitting and eruption of the filament acquired at 211 \AA\ can be observed in Supplementary Movie 4.

During the filament eruption, three dimmings appeared clearly in the 211 \AA\ images. The dimming regions are denoted by three colored boxes labeled 2, 3, 4 in Figure 4(e). Additionally, Figures 5(a)-(b) display running difference 211 \AA\ images at two different moments on March 9, 2019, highlighting the appearance of the dimming regions marked by the red, green, and blue arrows. In Figure 5(c), the red, green, and blue curves represent the normalized intensity variation of 211 \AA\ images from 12:00 UT to 15:00 UT in different partitions marked by the colored boxes 2, 3, 4 in Figure 4(e), respectively. The onset of the three dimming regions is indicated by the vertical dotted lines. The intensity of regions 2 and 3 decreased synchronously. Compared to the original intensity before the filament eruption at 12:00 UT, the intensity of regions 2, 3, and 4 decreased by about 80\%, 40\%, and 41\% at 12:36 UT, respectively. Usually, two dimmings emerge at the footpoints of filament or the S-shaped magnetic structures (Sterling, \& Hudson 1997). However, in this event, three dimming regions appeared during the filament eruption, which can be observed in Supplementary Movie 4. Dimming regions 2 and 4 are related to the eruption of the newly formed filament, while dimming regions 3 and 4 are linked to the eruption of the remaining part of the original filament after its splitting. Due to the long distance between stars and Earth, we cannot directly observe the stellar CMEs. Dimmings are suggested to be an indicator for the occurrence of CMEs on stars. In this case study, dimmings appeared during the eruptions of the newly formed filaments, while no CME was observed. To confirm whether a CME has occurred, in addition to dimmings, we also need other observational evidences as a supplement.

To illustrate the expansion of the bright loops and the eruption of the longer newly formed filament, two time-distance diagrams are made along the red line AB and the blue line CD marked in Figure 4(a) using 211 \AA\ images during the filament eruption from 12:00 UT to 12:30 UT on March 9, 2019, as shown in Figures 6(a)-(b). In Figure 6(a), the dashed lines marked by the numbers 1 and 2 outline the expansion of the bright loops and the eruption of the longer newly formed filament, respectively. Their projected velocities are estimated to be 8.44 km s$^{-1}$ and 44.89 km s$^{-1}$, respectively. In Figure 6(b), the dashed lines marked by the numbers 3 and 4 indicate the bidirectional jet that occurred during the magnetic reconnection between the filament and the surrounding loops. The velocities of the bidirectional jet are approximately 54 km s$^{-1}$ and 63 km s$^{-1}$, respectively. Note that the velocities are a low estimation, because they are estimations of the velocity component perpendicular to the line of sight. One jet was directed along the surrounding coronal loops, while the other was directed along the filament. Additionally, the dashed lines marked by the numbers 5 and 6 indicate the eruptions of the two newly formed filaments.

\subsection{Convergence motion and magnetic cancellation}

In Figures 7(a)-(d), the line-of-sight magnetograms observed by HMI on March 9, 2019, illustrate the distribution and temporal evolution of the magnetic fields. The white and black regions represent positive and negative polarities, respectively. The opposite fluxes continuously converge toward each other, leading to continuous flux cancellation. The blue arrows in Figures 7(b)-(d) indicate the sites of magnetic cancellation. Figure 7(e) displays the evolution of magnetic fluxes in the active region outlined by the red box in Figure 7(a). The black and blue lines represent the evolution of positive and negative magnetic fluxes in the selected region, respectively. The vertical black dotted line denotes the onset of the decreasing trend of both positive and negative fluxes. Overall, the trend of positive magnetic fluxes is the same as that of negative magnetic fluxes, suggesting a consistent evolution of the magnetic field polarities in the active region.

Figure 7(f) presents the time-distance diagram marked by the slit along the red line EF in Figure 7(b). We generated several time-distance diagrams along different orientations and selected the most illustrative one to demonstrate the movement of the eruptive structure and the convergence motion. The positive and negative polarities approached each other at speeds of 0.06 km s$^{-1}$ and 0.02 km s$^{-1}$, respectively. The convergence motion of two opposite polarities led to magnetic cancellation, which likely contributed to the magnetic structure formation of the longer newly formed filament. The conclusion is inferred from the following reasons: First, the convergence motion of the two opposite polarities preceded the magnetic cancellation; Second, the brightening occurred after the onset of the magnetic cancellation. Based on the observed causality, we suggest that the magnetic structure formation of the longer newly formed filament is a result of the magnetic cancellation driven by the convergence motion of two opposite polarities.

\subsection{Nonlinear force free field extrapolation}

To explore the filament's structure and elucidate the formation of dimmings during its eruption, we performed non-linear force-free field extrapolation based on the vector magnetic field observed by SDO/HMI. Figures 8(a) and 8(c) display the 211 \AA\ images at 11:00:09 UT and 12:02:21 UT on March 9, 2019, respectively. The black and blue dotted lines outline the surrounding loops at the east  of the filament, which could involve in the magnetic reconnection with the filament. The black arrows in Figures 8(a) and 8(c) indicate the filament. Figures 8(b) and 8(d) exhibit the magnetic structure of the filament from a top view and a side view, respectively, at 11:00 UT on March 9, 2019, obtained through non-linear force-free field extrapolation. The black and blue dotted lines in Figure 8(b) indicate the surrounding loops at the east of the filament at 11:00:09 UT and 12:02:21 UT, respectively. The filament's magnetic structure appears as a flux rope. The yellow lines denote the magnetic field lines originating from positive to negative magnetic fields near the leading negative sunspot of the active region, while the red lines denote the magnetic field lines from the following positive sunspot to the leading negative one. Two types of magnetic field lines compose the filament's magnetic structures. The white field lines depict the large-scale magnetic field of the active region. From the extrapolation, we observe three footpoints of the filament. The extrapolated magnetic structure of the filament exhibits a right hand twist, consistent with observations. Two dimmings observed in the 211 \AA\ wavelength correspond to the footpoints of the red magnetic field lines (see the green and blue boxes in Figure 4(e)). Another dimming (red box 2) corresponds to the eastward footpoint of the newly formed S-shaped filament outlined by the blue dotted lines in Figure 4. Typically, dimmings appear at filament's footpoints after its eruption. In this event, four dimmings should be observed since the original filament and the two newly formed filaments have four footpoints. Upon checking the 211 \AA\ observations, we find that the westward footpoint of the yellow magnetic field lines is found to be covered by the flare-loops (see Figure 4(f)).

\subsection{Schematic diagram of the filament evolution}

A schematic diagram (Figure 9) illustrates the change in the magnetic structures of the filament system before and after the filament eruption. The pink and black patches represent positive and negative polarities, respectively. The green and blue curved lines represent the magnetic fields before the filament eruption, while the orange curved lines depict the surrounding coronal loops. The red pentagrams indicate the position where the magnetic reconnection between the filament and the surrounding coronal loops occurs. Additionally, the red curved line represents the newly formed magnetic field line resulting from the magnetic reconnection. 

\section{Conclusion and Discussion} 

In this study, we investigated an interesting case study of a filament splitting and eruption associated with a B-class solar flare in the active region NOAA 12734 on March 9, 2019. By analyzing the EUV observations in the corona, we found that the filament splitting and subsequent eruption should be triggered by the magnetic reconnection between the filament and the surrounding magnetic loops. Prior to the filament splitting and eruption, the convergence motion and the magnetic cancellation occurred at the eastern footpoint of the filament, indicating that there is a slow magnetic reconnection occurring between the filament and the surrounding magnetic loops. The associated evidences of magnetic reconnection are also well observed, including obvious brightening at the junction of the filament and the surrounding magnetic loops, a bidirectional jet, and the formation of two newly formed filaments. During the eruption of the newly formed filaments, three dimmings were observed, corresponding to the three footpoints of the newly formed filaments. The appearance of these dimmings further suggests the occurrence of magnetic reconnection between the filament and the surrounding magnetic loops.

When the eastward part of the filament interacted with the surrounding magnetic loops, an evident brightening appeared and extended from the initial site to the surrounding loops. In the following, a large filament formed as a result of the magnetic reconnection. One footpoint of the newly formed filament was located at the eastward footpoint of the surrounding loops, while the other was situated at the footpoint of the original filament (the leading negative sunspot). Moreover, the remaining part of the filament gave rise to another filament. Subsequently, the two newly formed filaments erupted successively, accompanied by three noticeable dimmings. These observations confirm that the magnetic reconnection between the surrounding magnetic loops and the filament triggered the splitting of the original filament and the eruption of the newly formed filaments.

Magnetic cancellation was observed at the eastward footpoint of the filament and the westward footpoints of the surrounding magnetic loops approximately 19 hours before the onset of the filament eruption. This relatively long period prior to the filament instability  suggests the occurrence of slow magnetic reconnection between the filament's magnetic structure and the surrounding magnetic loops. Previous studies have highlighted the significant role of magnetic cancellation in the formation process of the eruptive magnetic structure (Chae et al. 2001; Yan et al. 2016; Yan et al. 2021). In this event, magnetic cancellation played a crucial role in separating one filament into two filaments, ultimately leading to their eruptions. This process is different from the ideal MHD processes of kink instability and torus instability. Instead, the magnetic reconnection between the filament and the surrounding magnetic loops, constituting a non-ideal MHD process, facilitated the observed phenomenon.

Yan et al. (2020a) observed an active-region filament that underwent reconnection with the surrounding magnetic loops, leading to the transfer of the filament materials from the filament to the overlying magnetic loops and resulting in the failed eruption of the filament. However, the event studied here differs significantly from that described by Yan et al. (2020a). In contrast to their findings, the magnetic reconnection in our event occurred between the filament and the surrounding magnetic loops, specifically those adjacent to the eastward end of the filament, rather than the overlying magnetic loops. This reconnection caused the original filament to split into two new filaments. These observations reveal a novel mechanism for the formation of double-deck filaments. Furthermore, the newly formed filaments erupted after their formation, accompanied by three clearly associated dimmings. Importantly, these eruptions were not associated with coronal mass ejections (CMEs), indicating that the appearance of a dimming does not necessarily imply the occurrence of a CME. Liu et al. (2012) proposed two possible scenarios for the formation of double-decker magnetic flux ropes prior to the onset of eruption, a process also supported by simulations conducted by Kliem et al. (2014). Our study suggests that the original filament can split into two newly formed filaments due to magnetic reconnection with the surrounding magnetic loops, providing additional evidence for the formation mechanism of double-decker magnetic flux ropes.

During the eruptions of the newly formed filaments, three distinct dimming regions located in the three footpoints of the two filaments. Two dimmings appeared at the footpoints of the original filament, while another dimming appeared at the eastward footpoints of the surrounding magnetic loops. To elucidate the relationship between the dimming regions and the two newly formed filaments, we performed non-linear force-free extrapolation based on the vector magnetograms observed by SDO/HMI. The NLFFF extrapolation illustrated that the magnetic structure of the original filament comprised two segments of magnetic field lines originating from different magnetic polarities. One segment connected the leading sunspot with the positive polarity, while the other linked the discrete negative polarity near the following sunspot with the following sunspot. The two dimmings were observed at footpoints of the magnetic flux rope represented in red (see Fig. 8b). This outcome suggests that the magnetic structure of the filament is more intricate than previously depicted in current models.

In this case study, there are several new and original elements obtained as follows: Firstly, the splitting and subsequent eruption of solar active region filaments presented in this paper is not usually observed. The magnetic reconnection between different magnetic structures resulting in the filament splitting can shed light on the explanation of double-deck filament formation. Secondly, the magnetic reconnection between the filament and the surrounding magnetic loops is clearly identified, including obvious brightening at their junction, jet-like ejection, bidirection flows, and subsequent filament splitting. Thirdly, the filament experienced a failed eruption, but the obvious dimmings were observed after the eruption. It implies that even though no CME occurred, the dimmings can be also observed during the filament eruption. The dimmings cannot be taken as the only signal to identify whether a CME has occurred on the Sun or Sun-like stars.

\acknowledgments

The authors thank the referee for constructive suggestions that helped greatly in improving the manuscript. We would like to thank the SDO/AIA teams for the high-cadence data support. This work is surpported by the Strategic Priority Research Program of the Chinese Academy of Sciences under grant No. XDB0560000, the National Science Foundation of China under grant No. 12325303, the Yunnan Fundamental Research Projects under grant No. 202301AT070347, and the Yunnan Key Laboratory of Solar Physics and Space Science (No. 202205AG070009).

\begin{figure}
\epsscale{.9}
\plotone{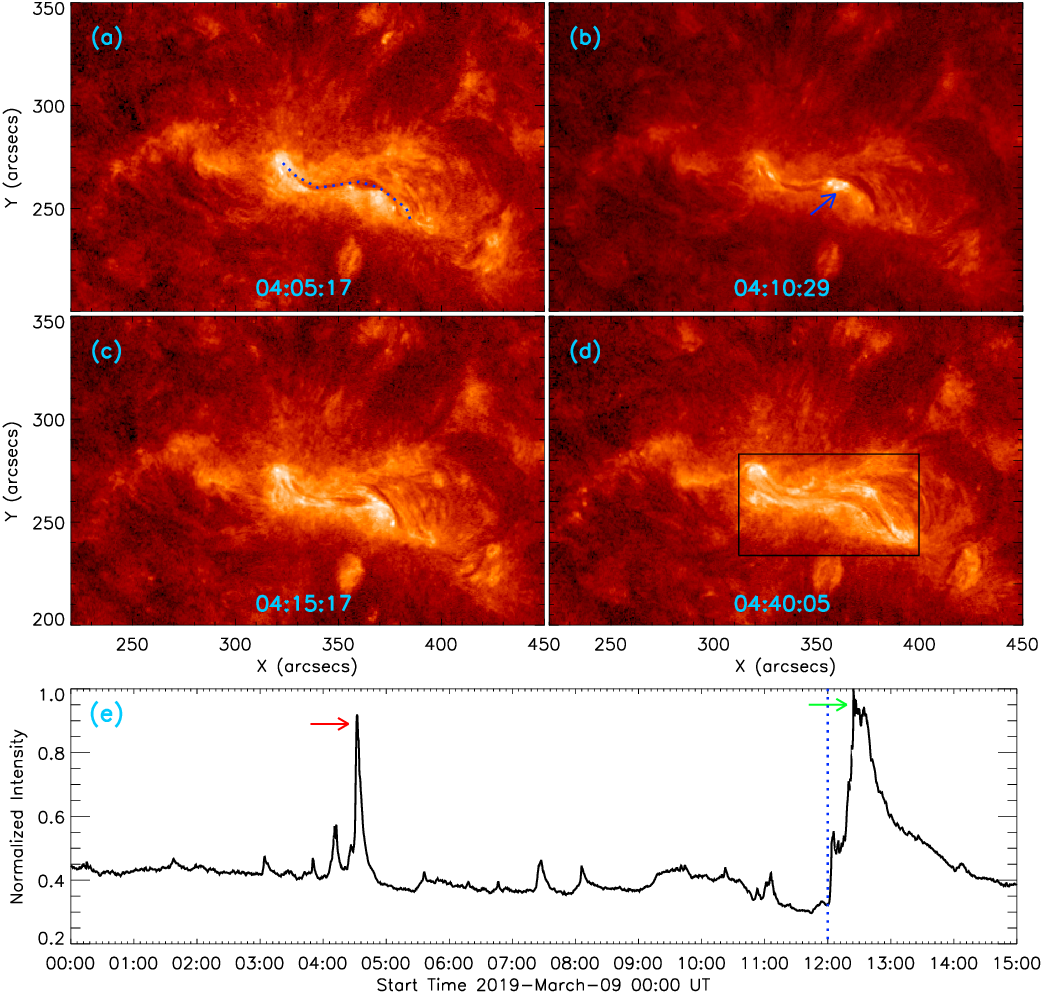}
\caption{(a)-(d): AIA 304 \AA\ images showing the evolution of the filament from 04:05:17 UT to 04:40:05 UT. The blue dotted line in panel (a) outlines the initial sigmoid structure of the filament. The blue arrow in panel (b) marks a brightening region, indicating the energy release. The activated filament is shown in panels (c) and (d). (e): Normalized intensity variation of the selected region marked by the black box in panel (d). The factor of normalization is the maximum value in the sub-field of views considered for the computation of the light curves. The red arrow points the higher peak of the intensity profile appearing during the first brightening. The green arrow indicates the brightening associated with the B6.1 flare. The vertical blue dotted line indicates the onset of the filament eruption. An animation of the 304 \AA\ images is available in Supplementary Movie 1. Its duration is 12 seconds, covering from 04:00 UT to 05:00 UT on March 09, 2019.
\label{fig1}}
\end{figure}

\begin{figure}
\epsscale{.9}
\plotone{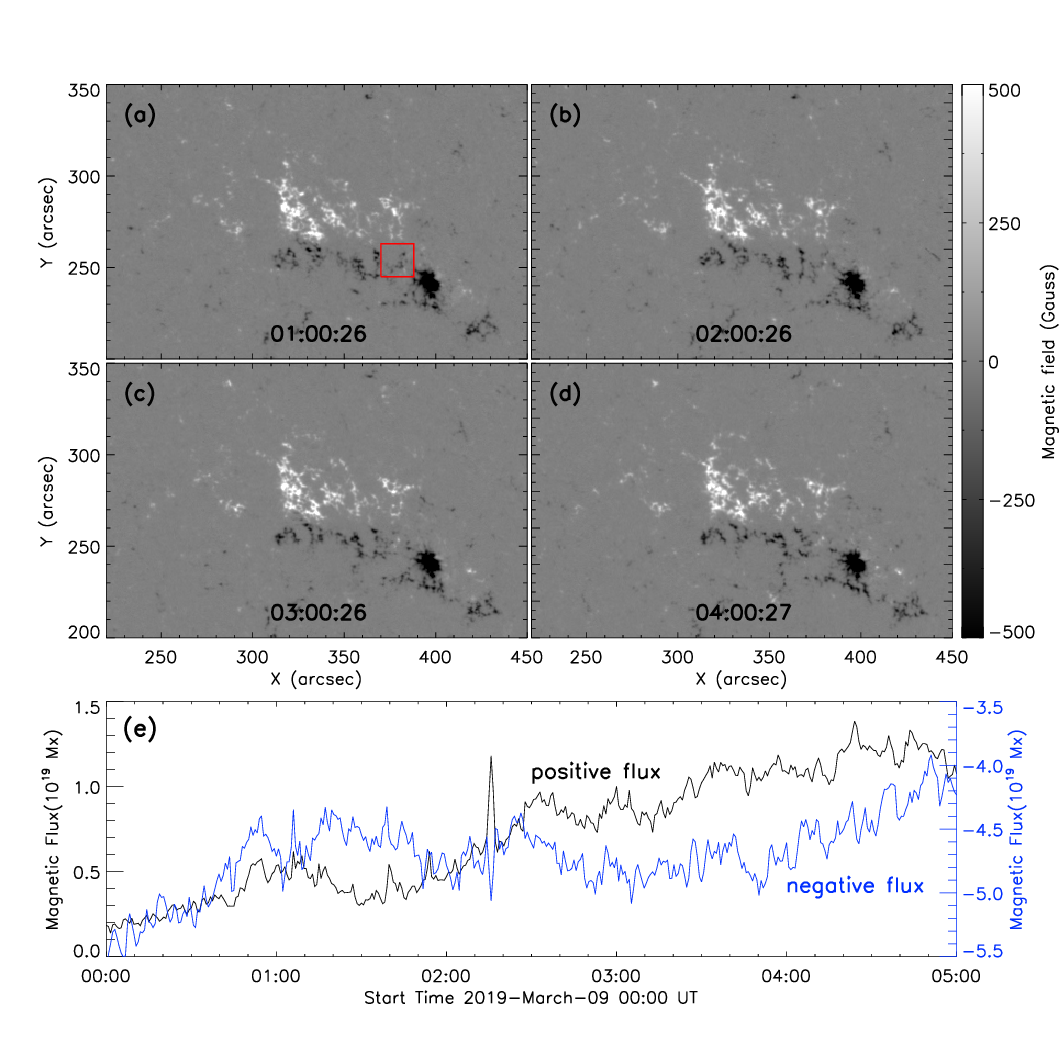}
\caption{(a)-(d): HMI line-of-sight magnetograms displaying the distribution of the magnetic fields with positive and negative polarities shown in white and black, respectively. (e): Temporal evolution of the positive (black line) and the negative (blue line) magnetic fluxes in the active region marked by the red box in panel (a). An animation of the magnetograms is available in Supplementary Movie 2. Its duration is 6 seconds, covering from 02:45 UT to 05:00 UT on March 09, 2019.
\label{fig1}}
\end{figure}

\begin{figure}
\epsscale{.9}
\plotone{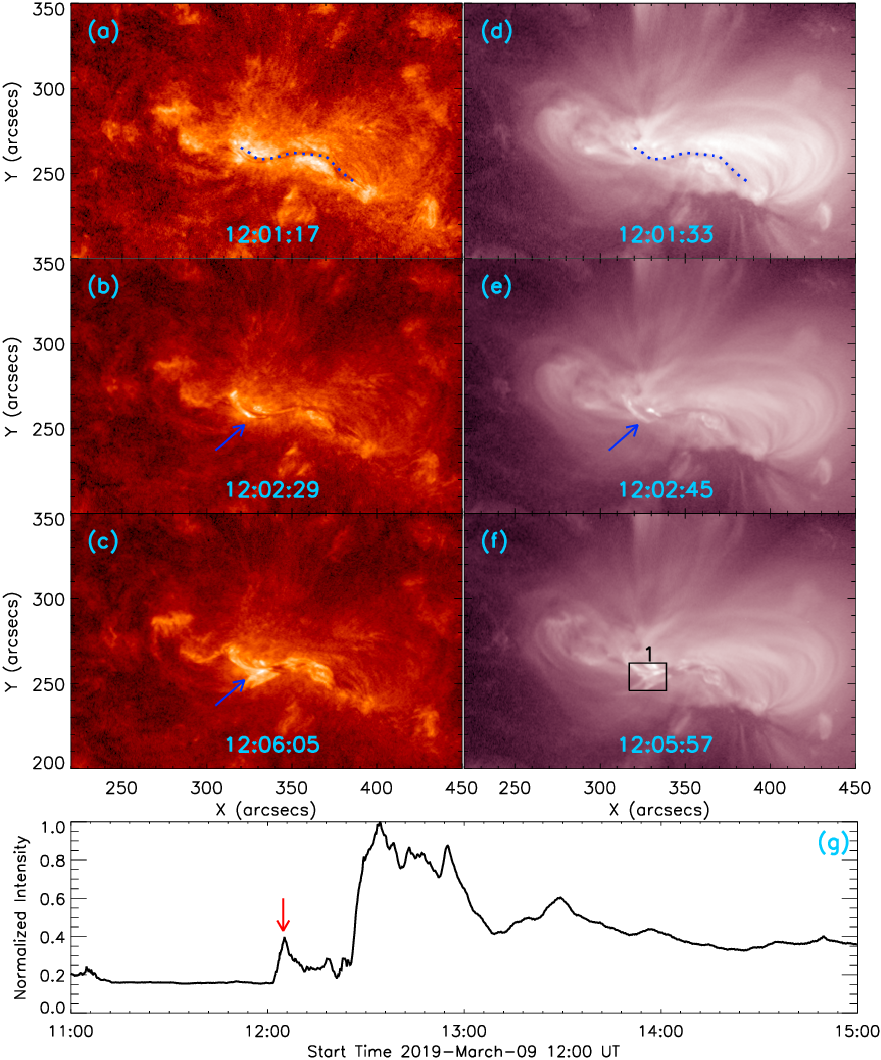}
\caption{(a)-(f): 304 \AA\ and 211 \AA\ images showing the eruption process of the filament. The blue dotted lines in panels (a) and (d) outline the initial structure of the filament. The blue arrows in panels (b), (c) and (e) mark the brightening region. (g): Normalized intensity variation of the selected segment marked by the black box in panel (f). The black curve shows the intensity variation normalized by its maximum value. The red arrow in panel (g) indicates the appearance of the brightening caused by the magnetic reconnection between the filament and the surrounding magnetic loops. An animation of the 304 \AA\ images is available in Supplementary Movie 3. Its duration is 6 seconds, covering from 12:00 UT to 12:30 UT on March 09, 2019. 
\label{fig1}}
\end{figure}

\begin{figure}
\epsscale{.8}
\plotone{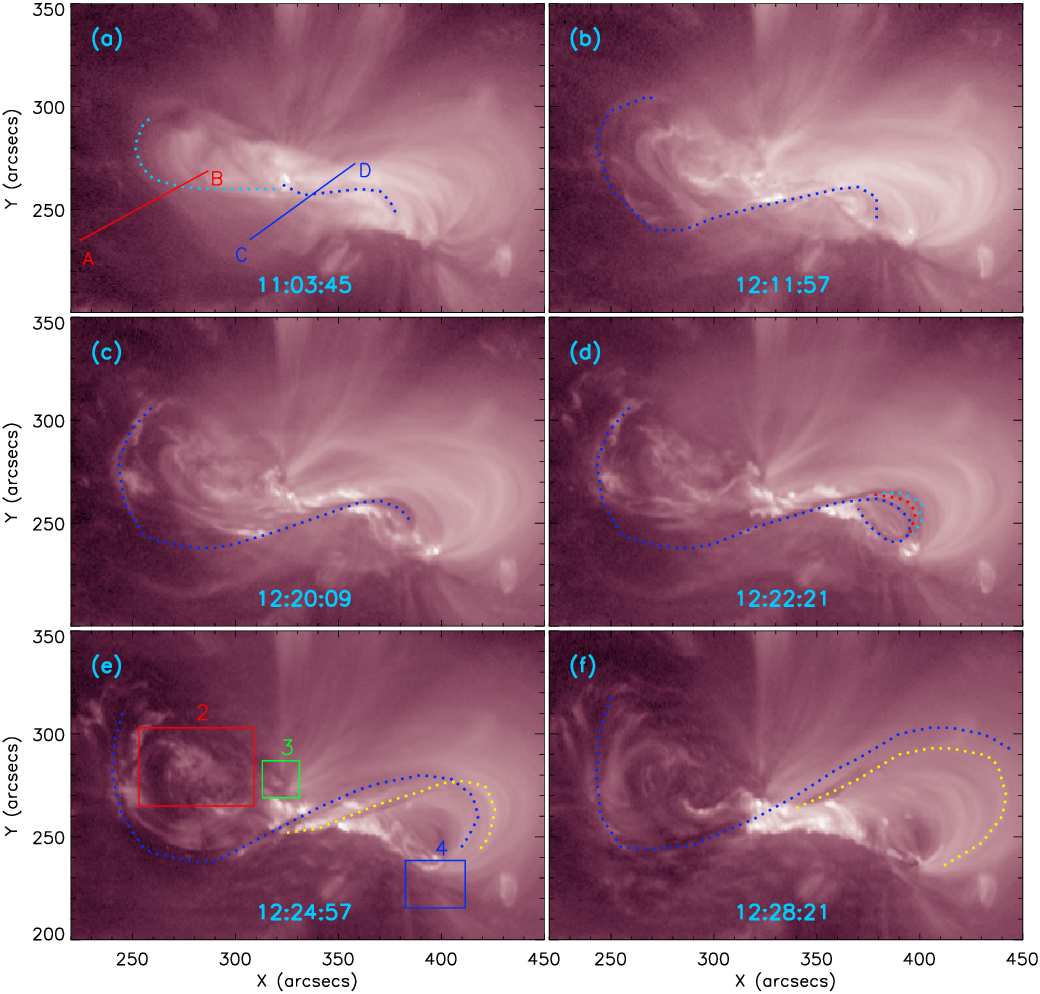}
\caption{(a)-(f): 211 \AA\ images showing the eruption process of the filament. The blue and the cyan dotted lines in panel (a) outline the shape of the filament before its eruption and the surrounding magnetic loops at the east of the filament. The blue and the yellow dotted lines in panels (b)-(f) indicates the two newly formed filaments. The lines AB and CD in panel (a) denote the position where the time slices in Figure 6 are made. Three segments of the 211 \AA\ image marked by the colored boxes 2, 3, 4 in panel (e) indicate three dimming regions. An animation of the 211 \AA\ images is available in Supplementary Movie 4. Its duration is 40 seconds, covering from 11:00 UT to 15:00 UT on March 09, 2019. 
\label{fig1}}
\end{figure}

\begin{figure}
\epsscale{1.}
\plotone{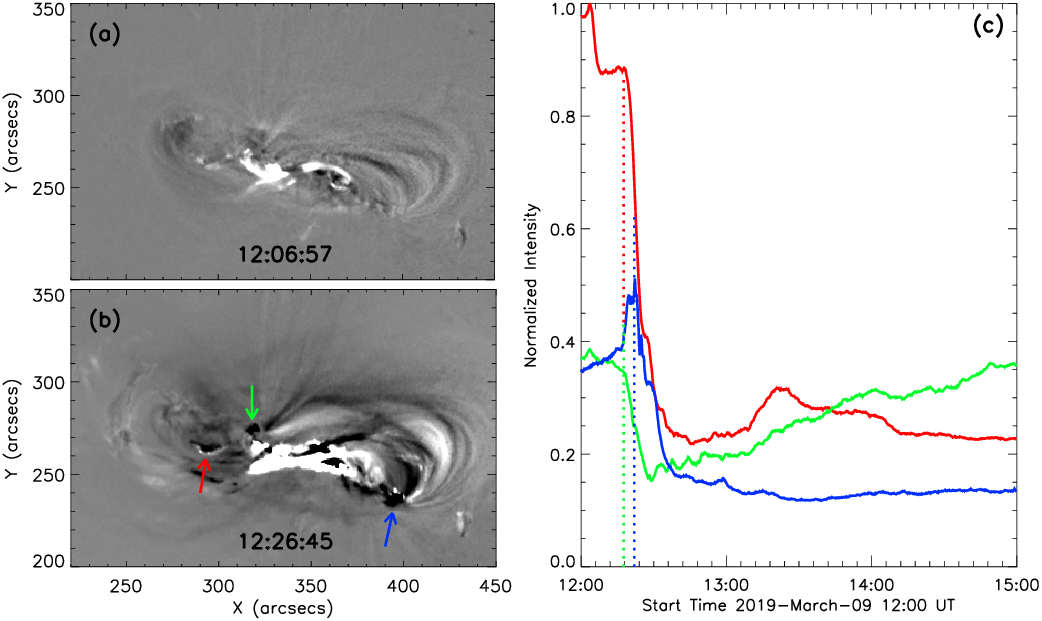}
\caption{(a)-(b): Running difference 211 \AA\ images at two different moments on 2019 March 09 to show the appearance of the dimming regions in the active region. The red, green, and blue arrows indicate the dimming regions. (c): Normalized intensity variation of 211 \AA\ images in the three selected segments. The red, green, and blue curves show the normalized intensity variation of 211 \AA\ images in different partitions marked by the colored boxes 2, 3, 4 in Figure 4(e), respectively. The start moments of the three dimming regions are indicated by the vertical dotted lines. 
\label{fig1}}
\end{figure}

\begin{figure}
\epsscale{.9}
\plotone{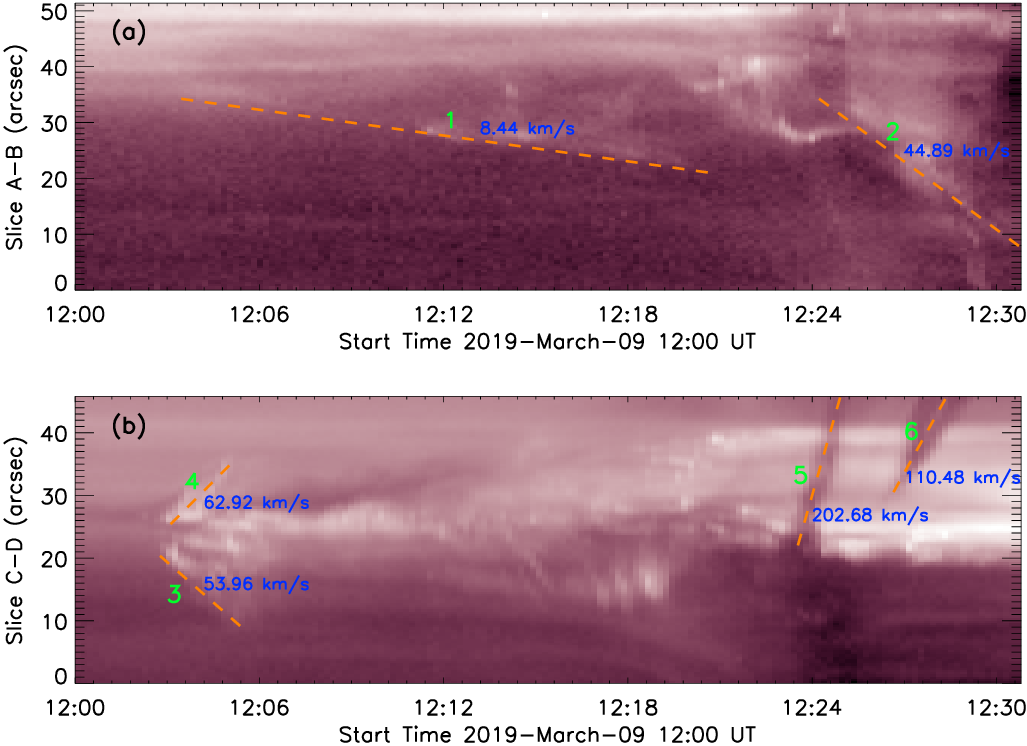}
\caption{(a)-(b): Time-distance diagrams from 211 \AA\ images along the red line AB and the blue line CD marked in Fig. 4(a) during the filament eruption from 12:00 UT to 12:30 UT on March 09, 2019. The dashed lines marked by 1 and 2 in panel (a) indicates the eruptions of the bright loops and the longer newly formed filament. The dashed lines marked by 3 and 4 in panel (b) outline the bidirection jet. The dashed lines marked by 5 and 6 indicate the eruptions of the two newly formed filaments. 
\label{fig1}}
\end{figure}

\begin{figure}
\epsscale{.9}
\plotone{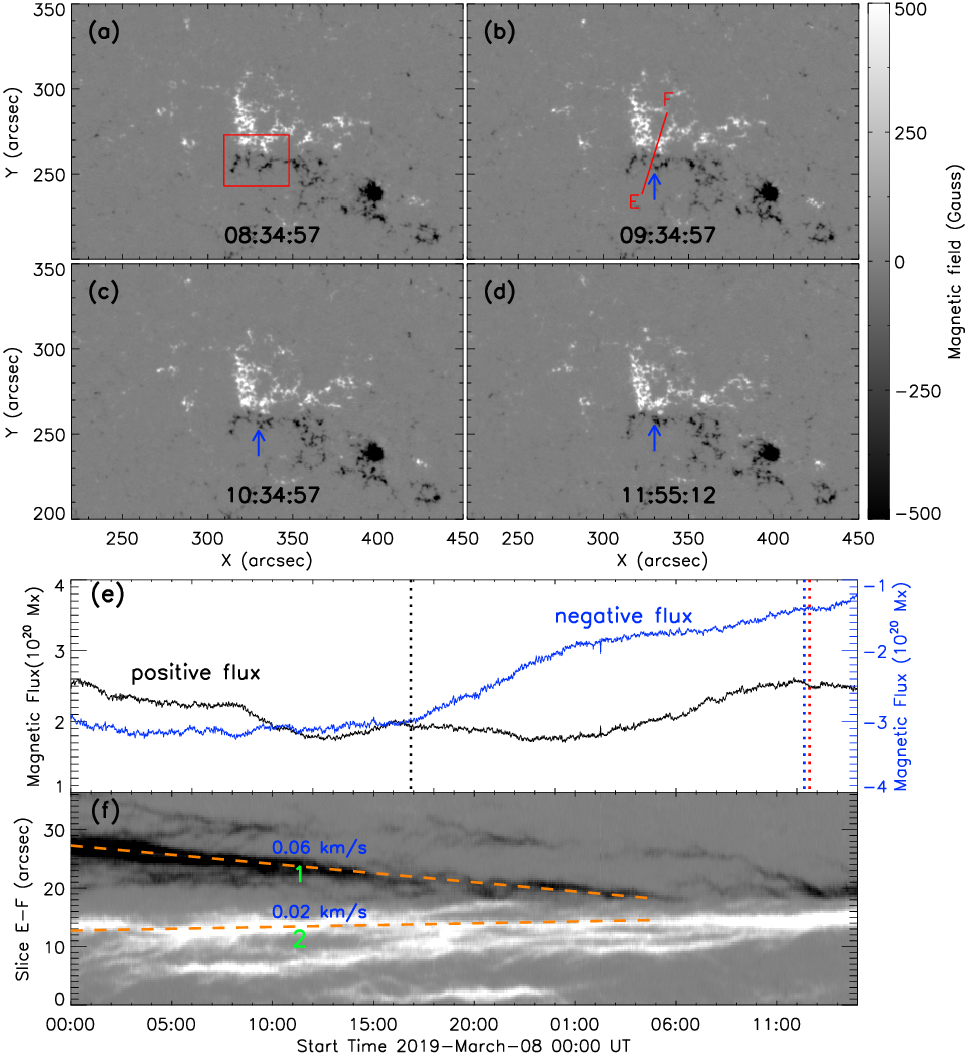}
\caption{(a)-(d): HMI line-of-sight magnetograms displaying the evolution of the magnetic fields with positive (negative) polarity shown in white (black). The blue arrows in panels (b)-(d) point to the sites of magnetic cancellation. (e): Temporal evolution of the positive (black) and the negative (blue) magnetic fluxes in the active region marked by the red box in panel (a). The three vertical dotted lines outline the beginning of the magnetic cancellation, the start and end moments of the associated B-class flare. (f): The time-distance diagram marked by slit along the red line EF in panel (b). 
\label{fig1}}
\end{figure}

\begin{figure}
\epsscale{1.}
\plotone{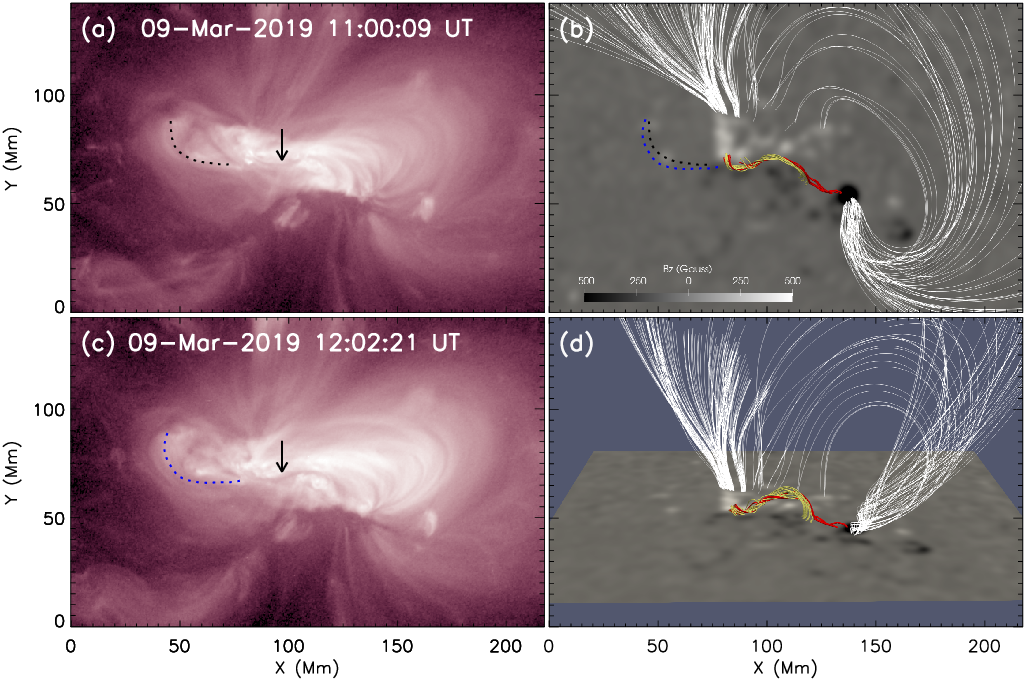}
\caption{The NLFFF extrapolation showing the magnetic structure of the filament. (a) and (c): The 211 \AA\ images at 11:00:09 UT and 12:02:21 UT on March 09, 2019. The black and the blue dotted lines indicate the surrounding loops at the east of the filament. The black arrows indicate the filament. (b) and (d): The NLFFF extrapolation showing the magnetic structure of the filament seen from top view and from side view at 11:00:09 UT. The black and the blue dotted lines in panel (b) indicate the surrounding loops at the east of the filament at 11:00:09 UT and 12:02:21 UT, respectively. The white lines show the large-scale magnetic loops of the active region.
\label{fig1}}
\end{figure}

\begin{figure}
\epsscale{.6}
\plotone{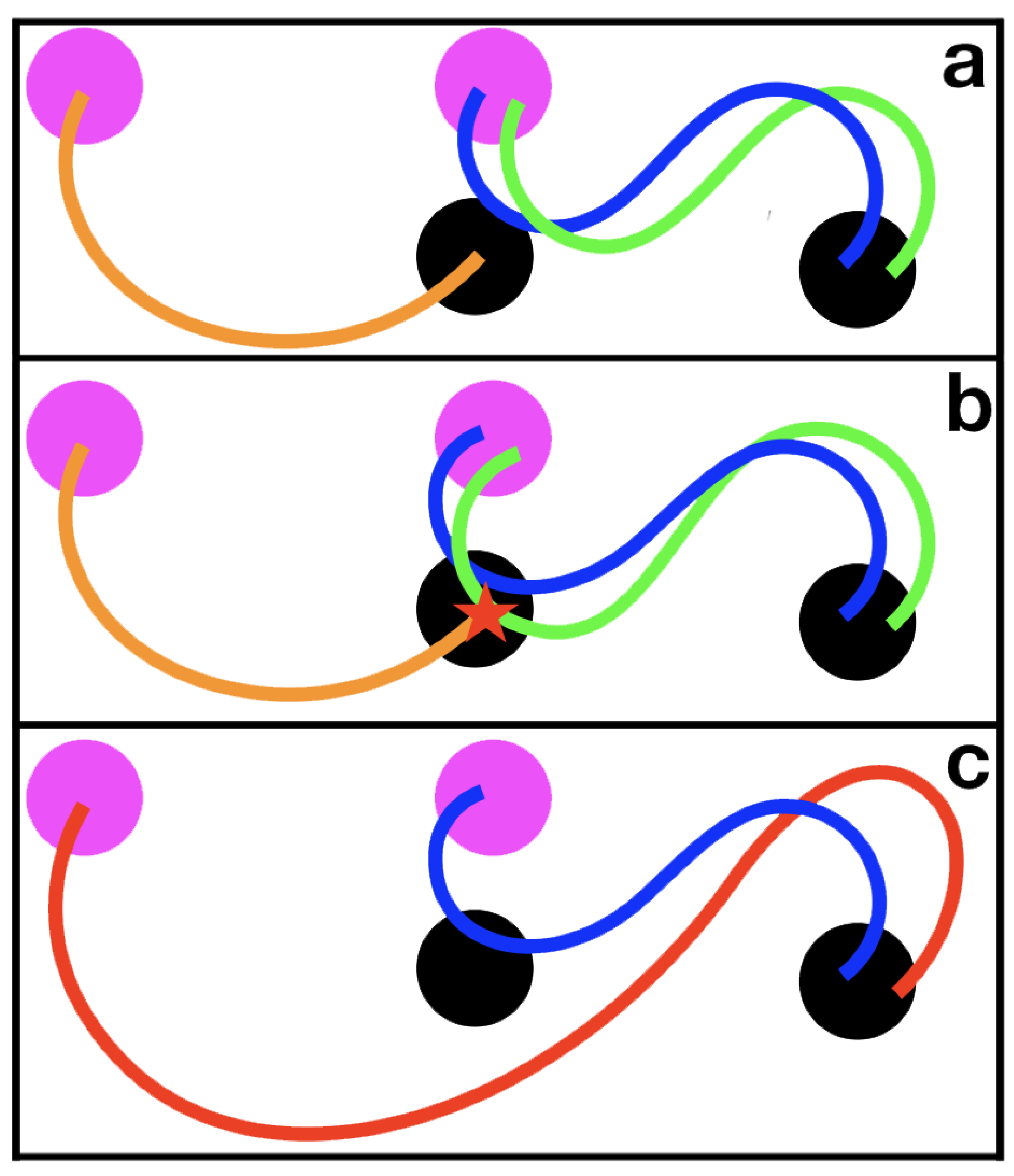}
\caption{A cartoon drawn to show the process of filament splitting due to the magnetic reconnection. The green and the blue lines show the twisted structure of the original filament. The orange line shows the surrounding magnetic loops. The pentagrams indicate the position where the magnetic reconnection occurs. The red line show the newly formed filament.
\label{fig1}}
\end{figure}

\end{document}